# Coherence of an Entangled Exciton-Photon State

A. J. Hudson[1,2], R. M. Stevenson[1], A. J. Bennett[1], R. J. Young[1],

C. A. Nicoll[2], P. Atkinson[2], K. Cooper[2], D. A. Ritchie[2] and A. J. Shields[1].

[1]Toshiba Research Europe Limited, 208 Cambridge Science Park, Cambridge CB4 0GZ, UK

[2]Cavendish Laboratory, University of Cambridge, Madingley Road, Cambridge CB3 0HE, UK

Abstract

We study the effect of the exciton fine-structure splitting on the polarisation-entanglement of photon pairs produced by the biexciton cascade in a single quantum dot. The entanglement is found to persist despite separations between the intermediate energy levels of up to 4μeV. Measurements demonstrate that entanglement of the photon pair is robust to the dephasing of the intermediate exciton state responsible for the first order coherence time of either single photon. We present a theoretical framework taking into account the effects of spin-scattering, background light and dephasing. We distinguish between the first-order coherence time, and a parameter which we measure for the first time and define as the cross-coherence time.

Semiconductor quantum dots are often referred to as 'artificial atoms' due to their three-dimensional electronic confinement and discrete energy levels [1]. One application for which they have been proved useful is the generation of polarisation-entangled photon pairs. In this Letter we study how the entanglement responds both to fluctuations in the dot's energetic fine-structure, and also to various dephasing mechanisms within the dot. We report for the first time on the distinction between different types of dephasing within the quantum dot, and investigate whether or not they can affect entanglement.

Entanglement arises when the wavefunction for two distinct entities can not be separated into the product of the wavefunctions of each entity. This leads to the counter-intuitive effect of a measurement on one body immediately affecting the wavefunction of the other, despite any spatial separation between the pair [2]. The most mature technique for the controlled generation of entanglement today is parametric down-conversion in non-linear crystals [3,4]. However many applications of quantum information processing, including efficient quantum computing [5] and scalable quantum communication [6], require an on-demand source of entanglement which parametric down-conversion is unable to provide due to its probabilistic nature [7].

Recently we demonstrated that triggered polarisation-entangled photon pairs may be generated from the radiative decay of a single quantum dot [8,9,10], as proposed by Benson *et al* [11]. The dot is initially excited to a zero-spin bound complex of two electrons and two heavy-holes (a biexciton.) From here it decays by the sequential emission of two photons, constrained to have total angular momentum zero, and returns to a ground (unexcited) state. If the intermediate states (excitons) of the dot are degenerate (fig. 1a) then the polarisations of the two photons are maximally entangled and the emission can be written as $(|H_{XX}H_X\rangle + |V_{XX}V_X\rangle)/\sqrt{2}$ where H (V) refers to the horizontal (vertical) polarisation and subscripts XX (X) denote the first (second) emitted photon.

However it is more usually the case that a fine-structure splitting (FSS) exists between the two orthogonal exciton states corresponding to the H and V polarised photons. This means that the decay cascade is forced to take one of two distinguishable paths (fig. 1b) and the photon polarisations are only classically correlated rather than entangled [12,13,14].

How small the FSS must be in order to cross over from polarisation-correlated to polarisation-entangled emission has not yet been addressed. By controlling the FSS of a single quantum dot with an in-plane magnetic field we are able to study the degree of entanglement of the emission as a continuously varying function of FSS. We develop a theoretical model for this function by considering the phase-evolution of the intermediate exciton states. We will show that entanglement is tolerant to small fluctuations in the FSS, and that it is possible to control the precise state of the output by manipulating the FSS.

Finally we measure the characteristic time of dephasing between superimposed H and V intermediate exciton-photon states, which we refer to as the cross-dephasing time. We find that this cross-dephasing time is significantly longer than the first order coherence time of either individual photon, making entanglement generation from the quantum dot biexciton cascade robust against environmental decoherence effects.

Our model and theoretical results derive from an analysis of the phase evolution of the decay cascade in the time-domain. Both the initial (biexciton) and final (ground) states of the dot are energy eigenstates and hence do not evolve over time. However once the dot has emitted the first photon, the intermediate state is a superposition of two exciton-photon eigenstates with energetic separation $S$ equal to the FSS, as the photon wavefunctions are both energy eigenstates and their energetic separation has no effect. The time, $t$, spent in this superposition state therefore introduces a phase difference between the two component states equal to $St/\hbar$. Accordingly the final 2-photon state is:

$$|\psi\rangle = \left(|H_{XX}H_X\rangle + \exp(iSt/\hbar)|V_{XX}V_X\rangle\right)/\sqrt{2}$$

from which a (pure state) density operator can be evaluated as $\hat{\rho}_{pure} = |\psi\rangle\langle\psi|$.

In order to construct the full density operator for the (mixed state) output it is then necessary to integrate this operator over all values of $t$. The emission of the exciton photon obeys Poissonian statistics and the probability of emission occurring between time $t$ and $t+dt$ in any given decay cycle is therefore equal to $\exp(-t/\tau_1)dt/\tau_1$, where $\tau_1$ is the radiative lifetime of the exciton. The density operator for the coherent light emitted from the dot is therefore

$$\hat{\rho} = \int_0^\infty \frac{1}{\tau_1} \exp(-\frac{t}{\tau_1}) \hat{\rho}_{pure} dt.$$

Other light which must be included in the complete density matrix for the emission can be divided into background light, spin-scattered light, and dephased light. Background light is uncorrelated in polarisation, originating from areas of the sample other than the dot. Spin-scattered light occurs when the spin of the exciton is scattered after emission of the first photon (which constitutes a measurement), collapsing the superposition and again leading to no polarisation-correlation on average.

With respect to dephased light, we highlight the conceptual difference between dephasing events that randomise the phase relationship of a single eigenstate with itself at an earlier time, and those that randomise the phase relationship between two superimposed eigenstates. We illustrate this difference in fig. 1c, which represents the wavefunction for the H and V polarised components of the superposition over time.

The first type of event occurs with characteristic time $\tau_2^*$ which is the pure dephasing time of the dot, and typically limits the first-order single-photon coherence time $\tau_2$ to 10-100ps [15,16]. Our experiments demonstrate however that such dephasing maintains the phase relationship between the two eigenstates, as shown schematically in fig. 1c.

The second type of event occurs less frequently, as illustrated by the third dephasing event in fig. 1c, and we assign to this the characteristic cross-dephasing time $\tau_{HV}$. Cross-dephasing not only randomises the phase of the exciton-photon wavefunction with respect to itself earlier in time, but crucially it randomises the phase relationship between the two superimposed eigenstates as shown. This could be due to a polarisation-dependent phonon interaction [17], or fluctuating fields from background charges. Such fields are well-known to cause spectral wandering of the emission energy but they predominantly affect each polarisation equally. However any anisotropic component could cause a cross-dephasing event. These events contribute to a reduction in the first-order cross-coherence, $g_{H,V}^{(1)}$, defined as the probability of a given photon pair from the dot being emitted with a well-preserved phase between the energy eigenstates. The time-averaged effect of spin-scattering is also to reduce the cross-coherence, but the two effects are distinct as any emission after a cross-dephasing event is still classically correlated in polarisation.

The overall density operator $\underline{\underline{\rho}}$ can be written in terms of $\tau_{HV}$, $\tau_{SS}$ and $\tau_1$ (the characteristic times of cross-dephasing, spin-scattering and the radiative recombination respectively, assumed constant with respect to the FSS.) In matrix form in the [$H_{XX}H_X$, $H_{XX}V_X$, $V_{XX}H_X$, $V_{XX}V_X$] basis this evaluates to be:

$$\underline{\underline{\rho}} = \frac{1}{4} \begin{pmatrix} 1 + kg'^{(1)}_{H,V} & 0 & 0 & 2kg^{(1)}_{H,V} z^* \\ 0 & 1 - kg'^{(1)}_{H,V} & 0 & 0 \\ 0 & 0 & 1 - kg'^{(1)}_{H,V} & 0 \\ 2kg^{(1)}_{H,V} z & 0 & 0 & 1 + kg'^{(1)}_{H,V} \end{pmatrix}$$

$$g'^{(1)}_{H,V} = \frac{1}{1+\tau_1/\tau_{SS}} ; g^{(1)}_{H,V} = \frac{1}{1+\tau_1/\tau_{SS}+\tau_1/\tau_{HV}} ; z = \frac{1+ix}{1+x^2} ; x = \frac{g^{(1)}_{H,V} S \tau_1}{\hbar}$$

Here $g'^{(1)}_{H,V}$ is the fraction of dot emission unaffected by spin-scattering, which is equivalent to the first-order cross-coherence $g^{(1)}_{H,V}$ in the absence of cross- dephasing. $k$ is the fraction of photon pairs that originate exclusively from the dot.

As ideally the emission is expected to vary from $(|H_{xx}H_x\rangle+|V_{xx}V_x\rangle)/\sqrt{2}$ at zero FSS to the classically correlated mixture of $|H_{xx}H_x\rangle$ and $|V_{xx}V_x\rangle$ states at large FSS, the figure of merit we select to characterise the degree of polarisation-entanglement from the dot is the fidelity of the output state with the $(|H_{xx}H_x\rangle+|V_{xx}V_x\rangle)/\sqrt{2}$ state, f. For the theoretical density matrix above this is given by;

$$f = \frac{1}{4}\left(1+kg'^{(1)}_{H,V}+\frac{2kg^{(1)}_{H,V}}{1+x^2}\right)$$

Inspection reveals a Lorentzian shape for the dependence of f on S, with width $\Delta S = 2\hbar/\tau_1 g^{(1)}_{H,V}$, a peak fidelity value of $f_{max}(S=0) = (1+kg'^{(1)}_{H,V}+2kg^{(1)}_{H,V})/4$, and base fidelity $f_{min}(S \to \infty) = (1+kg'^{(1)}_{H,V})/4$. In the limit of slow spin scattering, this agrees with numerical data points obtained in other work [18].

Figure 2a plots the fidelity as a function of the FSS. In the ideal case of no background light, spin-scattering or cross-dephasing, *f* varies from unity at *S*=0 for the maximally entangled state, to 0.5 for the entirely classically correlated state, as shown by the black line. Finite cross-dephasing is represented by the orange line, which demonstrates a reduction in $f_{max}$ but does not affect $f_{min}$. This is because in the limit of large FSS, the phase difference between the two eigenstates evolves so rapidly compared to the radiative lifetime that the time-averaged phase difference upon emission is zero regardless of the occurrence or not of cross-dephasing events. A similar effect is observed for finite spin-scattering (red line,) but here $f_{min}$ is also reduced as spin-scattering results in HV and VH polarised photon pairs added to the emission. The blue line represents finite background light, where $k=0.5$. The result is identical reduction in $f_{max}$ and $f_{min}$ compared to an equivalent amount of spin-scattered light, but the width is narrower, and equivalent to the ideal case.

The different widths are another interesting feature of fig. 2a, where the fidelity including finite cross-dephasing exceeds that of the ideal case for $S>1\mu eV$. This counter-intuitive result can be understood by plotting the phase of $z$ as a function of the FSS as shown in fig. 2b. As $S$ increases, the phase of $z$ tends from 0 to $\pi/2$, and the maximally entangled part of the emission is tending from $(|H_{xx}H_x\rangle+|V_{xx}V_x\rangle)/\sqrt{2}$ to $(|H_{xx}H_x\rangle+i|V_{xx}V_x\rangle)/\sqrt{2}$. The phase varies less rapidly with splitting for reduced first order cross-coherence, which allows more entanglement to persist in the $(|H_{xx}H_x\rangle+|V_{xx}V_x\rangle)/\sqrt{2}$ state for larger splitting.

This result also suggests that by manipulating the FSS we can control the phase of the output, a highly desirable outcome for many applications of entanglement generation in quantum computing. We note that this result is consistent with the observed phase of the emission in other experiments where entanglement is generated by filtering out the classical component of the output from a dot with large FSS [19].

In our experiments we used InAs self-assembled quantum dots, grown by molecular beam epitaxy and placed at the centre of a weak imbalanced $1\lambda$ cavity as reported previously [9]. Apertures of 2-3μm in diameter were etched into a metal shadow mask placed on the surface in order to isolate single dots. The dots were cooled to ~5K and excited non-resonantly by laser pulses at 80MHz. Emission was collected by a high numerical-aperture lens. The light was divided both spectrally (which distinguishes the first and second photons) and according to polarisation. Photons were counted by avalanche photodiodes (APDs) and coincidences per excitation cycle between the biexciton and exciton photons were recorded, as described previously [9].

The degree of correlation, $c_\mu$, is defined (for an unpolarised system) in any given polarisation basis μ by

$$c_\mu = \frac{g_{xx,x} - g_{xx,\bar{x}}}{g_{xx,x} + g_{xx,\bar{x}}}$$

where $g_{xx,x}$ and $g_{xx,\bar{x}}$ are the second order correlation functions in the μ basis when detecting the XX photon with a co-polarised and cross-polarised X photon respectively. This degree of correlation is shown in fig. 3 (inset) for a dot 'A' measured in 3 bases: rectilinear, diagonal and circular. From these three measurements alone it is possible to calculate the fidelity of the output |ψ> with the maximally entangled $(|H_{XX}H_X\rangle + |V_{XX}V_X\rangle)/\sqrt{2}$ state via the simple equation

$$f = (1 + c_{rectilinear} + c_{diagonal} - c_{circular})/4.$$

This result is valid for any unpolarised source (verified experimentally within 1.8% error for these dots) which greatly simplifies the experiment compared to measuring the full density matrix of the state.

Data from two dots is shown in fig. 3. Both dots had approximately zero FSS in the absence of magnetic field and had peak fidelities f = 0.75±0.04 and f = 0.74±0.01, far in excess of the classical limit (0.5.) Magnetic fields up to 4T were applied in order to increase the FSS [20], and at each point both the fidelity and the FSS were measured. The FSS is determined directly from lorentzian fits to exciton and biexciton photoluminescence with a precision of 0.5μeV, as described previously [21].

Lorentzian fits are shown in fig. 3. and match the shape of the curves excellently, with full-width at half maximum values of 3.3±0.4μeV and 4.2±0.8μeV. As expected f drops below 0.5 for large FSS as the photons tend towards entirely classical polarisation-correlation mixed with uncorrelated light.

The lifetime of the exciton was found from the time-resolved photoluminescence spectrum (fig. 4a.) The intensity decays exponentially with time and the lifetime is determined from a best fit to this decay. The exciton lifetime was measured as $\tau_1$ = (891±11)ps and (881±35)ps respectively for dots A and B. The lifetimes are approximately independent of magnetic field, as shown for dot A in the inset of fig. 4a. A lifetime of $\tau_1 \sim$ 900ps (equivalent to a natural linewidth of ~0.7µeV) predicts a linewidth to the fidelity vs. FSS curves of ≥1.5µeV, which is of the same order as observed experimentally.

The presence of any cross-dephasing in the system would uniquely reduce f at *S*=0, without affecting f for large *S*. Based on the actual fidelity measurements at zero and maximum splitting, the corresponding reduction in $f_{max}$ compared to the value predicted by $f_{min}$ is 0.06±0.09 and 0.08±0.06 for dots A and B respectively, which does not show strong significance. This agrees with previous measurements where visibility of biphoton interference showed no signs of reduction through dephasing [10]. Determined instead from the Lorentzian fits, the fidelity reductions are 0.11±0.03 and 0.08±0.07 which equates to an average cross-dephasing time >2 ns. This is significantly more than the radiative lifetime, which suggests cross-dephasing is weak. We note that a more precise way to quantify the effect of cross-dephasing is to look for a discernibly weaker correlation in the diagonal and circular bases at zero FSS than in the rectilinear (H-V) basis. No difference is detectable from the correlations in fig. 3 (inset).

For comparison, the single-photon coherence time $\tau_2$ was measured with a first order interference experiment using a Michelson interferometer [22]. One arm of the interferometer was delayed relative to the other and the amplitude of the interference pattern recorded as a function of the delay (fig. 4b.) The rate of decay with delay time of the visibility of the interference fringes defines the first order coherence time. The measured coherence times for dots A and B were $\tau_2$ = (88±7)ps and

(110±3)ps, using the same excitation conditions (power, wavelength) as for the correlation measurements. The measured coherence times are short enough compared to the radiative lifetime that they are the same with error as the pure dephasing time $\tau_2^*$. Thus the measured cross-dephasing time is at least an order magnitude longer than the exciton pure dephasing time in the quantum dots. We note that the exponential decay of visibility with delay implies a Lorentzian lineshape for the exciton state, which in turn signifies that inhomogeneous spectral broadening does not affect the emission significantly.

In conclusion we have given a theoretical framework for the emission of polarisation-entangled photons from a quantum dot, noting especially the distinction between exciton dephasing and the previously unconsidered cross-dephasing that can affect entanglement. Within the accuracy of our experiment, we find that dephasing does not strongly affect entanglement. Additionally the tolerance of the degree of entanglement to small fluctuations in the FSS has been demonstrated. This is of great importance to any potential large-scale implementations of this entanglement-generation scheme, as variation in the FSS from dot to dot is inevitable [21]. It also suggests that manipulation of the FSS offers a way of controlling the entangled state of the emitted biphoton.

The authors gratefully acknowledge financial support from QIP IRC, EC FP6 Network of Excellence, Sandie and the EPSRC.

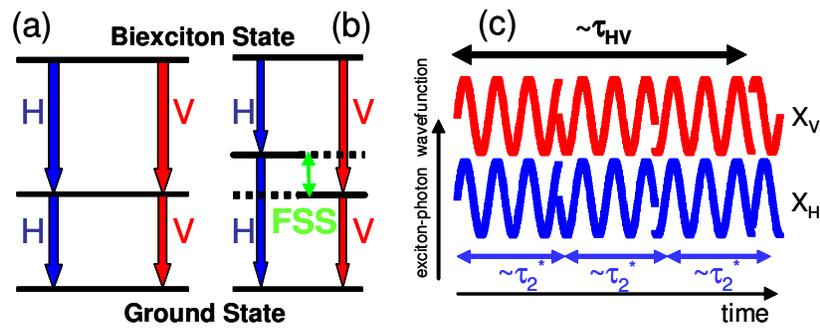

Fig. 1 (a) decay paths in a degenerate quantum dot; (b) decay paths in a non-degenerate (split) quantum dot; (c) wavefunction evolution of the superimposed intermediate states showing 3 dephasing events. The first 2 are single-photon decoherence events and do not affect the phase relationship of one field relative to the other. The 3rd event is a cross-coherence dephasing event and randomises the relative phase.

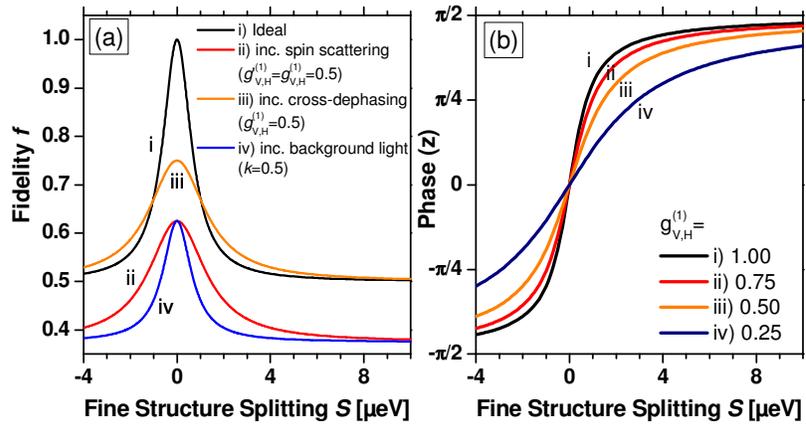

Fig. 2. (a) Predicted fidelity as a function of FSS for a quantum dot. Ideal behaviour is shown as a black line, and the effects of only spin-scattering, dephasing, or background light are shown by red, orange and blue. (b) Corresponding phase of z. Different curves show effect of reducing first-order cross coherence as indicated.

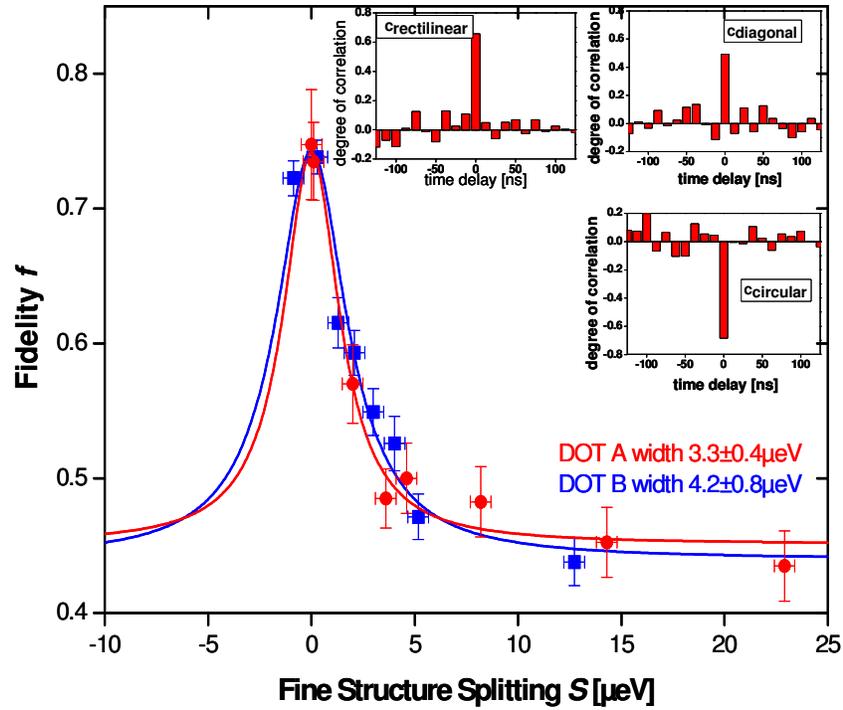

Fig. 3. Fidelity of the output state with the maximally entangled state as a function of fine-structure splitting, measured for two typical dots A (red discs) and B (blue squares). Solid lines are lorentzian fits to the data points. Inset: individual correlations for dot 'A' in the rectilinear, diagonal and circular bases.

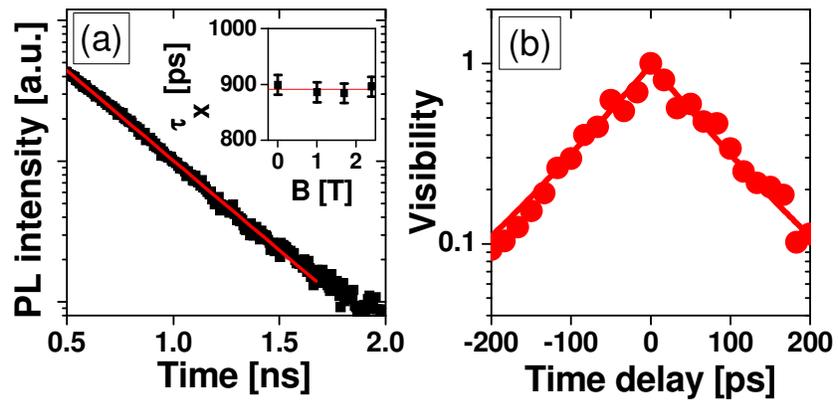

Fig. 4. (a) Exponential decay of exciton photoluminescence intensity with time for dot 'A'. Red line indicates fit used to determine lifetimes. Inset shows exciton lifetime with varying magnetic field. (b) Visibility of single photon interference fringes for dot 'A', used to determine first-order photon coherence time.